\documentclass[12pt,preprint]{aastex}
\newcommand{\eqb}{\begin{eqnarray}}
\newcommand{\eqe}{\end{eqnarray}}
\newcommand{\melec}{m_{\rm e}}
\newcommand{\tesc}{t_{\rm esc}}
\newcommand{\sigmaT}{\sigma_{\rm T}}
\newcommand{\nubreak}{\nu_{\rm bf}}           
\newcommand{\nusync}{\nu_{{\rm s,}18}}
\newcommand{\nucompt}{\nu_{{\rm c,}27}}

\newcommand{\elcomp}{\ell_{\rm e}}
\newcommand{\lowq}{q_{\rm e}}
\newcommand{\gammamax}{\gamma_{\rm max}}

\newcommand{\elinject}{Q_{\rm e}}
\newcommand{\tcross}{t_{\rm cr}}
\newcommand{\tvar}{t_{\rm var}}

\newcommand{\tvarthree}{t_3}
\newcommand{\doppler}{\delta}

\newcommand{\microm}{\,\mu\/{\rm m}}

\slugcomment{to appear in ApJ, 2002}

\shorttitle{Modelling the TeV emission of Blazars}
\shortauthors{Konopelko et al.}

\begin{document}


\title{Modelling the TeV
$\gamma$-ray spectra of two low redshift AGNs: Mkn~501 \& Mkn~421}


\author{A. Konopelko}
\affil{Max-Planck-Institut f\"ur Kernphysik, Postfach 10 39 80,
  Heidelberg, Germany}
%\email{konopelk@mickey.mpi-hd.mpg.de}
\author{A. Mastichiadis}
\affil{Department of Physics, University of Athens, Panepistimiopolis,
  GR 15783, Zografos, Greece}
\author{J. Kirk}
\affil{Max-Planck-Institut f\"ur Kernphysik, Postfach 10 39 80,
  Heidelberg, Germany}
\author{O.C. de Jager}
\affil{Unit for Space Physics, Potchefstroom University, 2520, South
  Africa}
\author{F.W. Stecker}
\affil{Laboratory of High Energy Astrophysics, NASA Goddard Space
  Flight Center, Greenbelt, MD 20771, USA}



\begin{abstract}
We discuss the results of modelling the
TeV $\gamma$-ray spectra
of two AGNs, Mkn~501 and Mkn~421 that have almost
the same redshifts: $z=0.031$ and $z=0.034$,
respectively. The effect
of intergalactic
$\gamma$-ray absorption is treated as an uncertainty in the
measurement of the intrinsic spectrum.
Although the objects differ, we obtain satisfactory fits for both of
them in a synchrotron self-Compton scenario.
Compared to previous models, our fits are characterised by higher
values of the Doppler factor ($\doppler\geq 50$)
and an electron injection spectrum extending to higher
energies ($\gammamax \geq 1.5 \times 10^{5}$).
In the case of Mkn~421, the observed difference in
spectral slope in X-rays and TeV $\gamma$-rays
between the high and
low states can be explained as a variation of a single parameter ---
the maximum energy $\gammamax mc^2$ at which electrons are injected.
\end{abstract}




\section{Introduction}

Ground-based detectors, utilizing the imaging
atmospheric \v{C}erenkov technique, provide an effective tool for the
study of
cosmic TeV $\gamma$-rays. Recently, several celestial objects
have been identified as TeV $\gamma$-ray emitters using this technique
\citep{ong98,cataneseweekes99}.
Among these are a few active galactic nuclei located at
different redshifts: Mkn~421 ($z=0.031$) \citep{punchetal92},
Mkn~501 ($z=0.034$) \citep{quinnetal96}, 1ES~2344+514 ($z=0.44$)
\citep{cataneseetal98}, 1ES~1959+650 ($z=0.048$) \citep{nishiyamaetal00},
PKS~2155-304 ($z=0.117$) \citep{chadwicketal99} and 1ES~1426+428 ($z=0.129$)
\citep{horanetal02}.

Several explanations have been advanced for the formation of the
spectrum of these objects, starting from an injection of energy in the form of
either energetic hadrons
\citep{mannheimandbiermann92,mannheim93,mannheim98,mueckeprotheroe01,
aharonian00} or energetic electrons
(e.g., \citet{maraschighisellinicelotti92,marschertravis96,inouetakahara96,
kinotakaharakusunose02}). At present it is not possible to eliminate either
of these possibilities on the basis of observation. However, the most
widely investigated model involves the injection of relativistic
electrons and the production of TeV photons via the synchrotron
self-Compton (SSC) mechanism
and in this paper we adhere to this scenario.

Some forty years ago it was suggested that
TeV $\gamma$-rays could be absorbed
by interacting with diffuse interstellar or intergalactic
infra-red (IR) radiation \citep{nikishov62,gouldschreder66}.
Following the discovery by the {\it EGRET} team that the blazar 3C279
at a redshift of 0.54 was an emitter of a powerful flux of high energy
$\gamma$-rays \citep{hartmanetal92}, it was suggested by 
\citet{steckerdejagersalamon92} 
that such powerful extragalactic sources
would be useful as probes of the extragalacic infrared radiation 
because of the absorption caused by pair-production interactions of their 
TeV $\gamma$-rays. \citet{steckerdejagersalamon92} further derived the 
formula for
the redshift dependence of this absorption and showed that the absorption
coefficient would be highly dependent on redshift. 

Owing to the lack of direct measurements
of the IR background radiation in the wavelength range
$1$--$50\,\microm$ \citep{hauserdwek01}, a computation of
the opacity of the intergalactic medium to TeV
$\gamma$-rays must be based on a model
of the diffuse radiation field.
The lower limits on the
extra-galactic background light (EBL) at $7\microm$ and $15\microm$
set by ISOCAM measurements \citep{franceschinietal01}
have ruled out a number of such models that
underestimate the EBL distribution and have
significantly constrained others. Improved computations of the intergalactic
opacity reveal, in turn, new information on the intrinsic TeV
spectra of AGN. In this paper we apply the SSC model to the two best
observed TeV sources, using opacities computed from models of the EBL
that satisfy the most recent constraints.

In principle, one can attempt to predict directly the TeV measurements,
by combining
a theoretical model of intrinsic spectrum of the AGN with one of the
intergalactic absorption. This would enable an unambiguous comparison of
prediction with observation and, via a maximum likelihood approach, an
estimate of the \lq\lq best-fit\rq\rq\ parameters. However, the absorption
model is subject to observational constraints independent of the TeV
measurements. These constraints allow a range of possibilities,
each of which can be associated with
a different likelihood of realisation. In order to take this into
account, it would be necessary to assign a range of values to each predicted
flux point, confusing the procedure of maximising the overall likelihood.

In this paper, we adopt a different approach, similar to that used in
\citet{konopelkoetal99}. There, it was
shown that for the EBL distribution given
by \citet{malkanstecker98} the
homogeneous SSC model provides a reasonable fit
to the intrinsic TeV $\gamma$-ray spectrum of Mkn~501.
Here, we use the revised distributions of the EBL given by
\citet{malkanstecker01} to estimate both the magnitude and
the uncertainty in the absorption and
construct intrinsic spectra using several sets of TeV data, before
fitting them with a homogeneous SSC model.
In particular, we use the results
recently obtained by the Whipple \citep{krennrichetal01} 
and HEGRA \citep{aharonianetal02} groups on Mkn~421
during its flaring state in 2001, which provide
statistics unprecedented in TeV astronomy, with more than $30,000$ photons
detected.

Previous discussions of the intrinsic spectrum of TeV blazars have
used specific forms for the absorption optical depth, rather than
estimating an uncertainty. Thus,
\citet{konopelkoetal99} obtained the intrinsic spectrum of Mkn~501 
using the combined
spectrum of the Whipple and HEGRA groups and applying the optical 
depth, $\tau =\tau (E_\gamma, z)$,
calculated by \citet{dejagerstecker98} based on the predictions 
on SED of intergalactic
IR photon field from \citet{malkanstecker98}. The unfolded spectrum was close
to a power-law of index 2.0 within the energy range starting from 500~GeV 
and extending up to 20~TeV.
Using a similar method for the optical depth derived from simplified
calculations of the SED
of EBL over optical, IR and far-IR wavelengths, 
\citet{kneiskemannheimhartmann02} obtained an
intrinsic spectrum of Mkn~501, which is consistent with a power-law of 2.49.
Using
another similar method, \citet{dejagerstecker02} reconstructed the 
intrinsic spectrum of
Mkn~501 as a power-law of index $\sim 1.6$--$1.7$ up to $\sim 5-10\,$TeV.
\citet{protheroemeyer00} had previously obtained an upturn in the 
intrinsic spectrum in this
energy range. However,
such an upturn does not occur when the recalibrated HEGRA
data for Mkn~501 \citep{aharonianetal01} and a
more realistic SED of the EBL is used, in which the
$60\microm$ flux reported by \citet{finkbeinerdavisschlegel00} is 
not taken into
account. \cite{aharonianetal02a} have also
derived the intrinsic spectrum of Mkn~501 for a specific SED of EBL
which also excludes the controversial $60\microm$ data point.

Having constructed the intrinsic spectrum we use the SSC model described
by \citet{mastichiadiskirk97} to obtain a fit. Several groups have
recently published fits to
both the X-ray and TeV $\gamma$-ray spectra of Mkn~421 and Mkn~501
(see e.g., \citet{tavecchioetal01,krawczynskietal01,kinotakaharakusunose02}) 
using
schemes similar to this one. However in these papers,
the effect of intergalactic
IR $\gamma$-ray absorption was not emphasized.
Here we test the SSC
model against the {\em de-absorbed} TeV $\gamma$-ray spectra of two BL~Lac's, 
Mkn~421 and
Mkn~501. The IR absorption of multi-TeV $\gamma$-rays substantially 
changes the intrinsic
source spectra and strongly influences
the optimum values of the model parameters we find.
In particular, taking account of extragalactic absorption moves the peak
of the intrinsic spectrum to higher energies. Thus, \citet{tavecchioetal98}, 
by not taking account of absorption, placed the Compton
peak in the Mkn 501 in the sub-TeV range and could not generate a
logically consistent SSC model, whereas \citet{dejagerstecker02}, 
taking account of absorption, showed that the peak in the 
intrinsic spectrum of Mkn 501 was $\sim$ 8 TeV.

The BL~Lac objects Mkn~501 and Mkn~421 have similar redshifts ($z=0.031$
and $z=0.034$, respectively) so that the effects of intergalactic absorption
are almost identical.
Observations of both of these objects show a softening of the spectrum
to higher energies. However, the intrinsic, {\it de-absorbed}
spectra, are very different, while the two objects have rather
different variability patterns.
Nevertheless, the SSC model provides a reasonable fit to both of them, 
indicating large
Doppler factors ($\doppler>50$) and high, but different, values for the maximum
energy at which electrons are injected. Thus, the exponential cut-offs
found at similar energies in observations of
both sources should be interpreted as due to the effect of intergalactic
absorption and not taken as evidence of
similarities in the physical conditions intrinsic to the two sources.
Recent data for other AGNs detected at larger redshifts ---
1ES1426+428 ($z= 0.129$) and PKS 2155-304 ($z=0.117$), are consistent
with this interpretation.

\section{Observations}

Since their discovery in TeV $\gamma$-rays two BL Lac objects Mkn~501 and 
Mkn~421
have been continuously monitored by a number of ground-based imaging 
atmospheric
\v{C}erenkov telescopes, e.g., CAT \citep{barrauetal98}, HEGRA 
\citep{konopelkoetal99a}, 
Whipple \citep{finleyetal01}. Even though the calibration of a \v{C}erenkov
telescope is not an easy procedure, it can ultimately be tested by joint 
measurements
of the energy spectrum of a source selected as a standard.
Recent measurements of the TeV energy spectrum of the Crab Nebula
performed by these groups are consistent within the quoted statistical 
errors (see 
\citet{aharonianetal00}). Thus spectral measurements performed on any 
particular
source by different groups can be considered
as complementary. In this section we briefly review the results of 
observations
of several TeV-emitting BL~Lac objects in X-rays and $\gamma$-rays.

\subsection{Mkn~501}
The BL~Lac object Mkn~501 was first detected by the Whipple group 
\citep{quinnetal96} in
TeV $\gamma$-rays and was later confirmed by the HEGRA group 
\citep{bradburyetal97}.
Early detections of Mkn~501 revealed a very low flux of TeV $\gamma$-rays,
at the level of about 0.5~Crab (1 Crab corresponds to the constant flux from
the Crab Nebula). However, in 1997 Mkn~501 exhibited an unprecedented
flare in TeV $\gamma$-rays with an integral flux of up to 10~Crab. A very 
long exposure on
this source, lasting almost 6~months, yielded unprecedented statistics for TeV
$\gamma$-rays, which provided very accurate measurements of the spectrum 
\citep{samuelsonetal98,krennrichetal99,
aharonianetal99,aharonianetal01,djannatiataietal99}
over the energy range from 250~GeV up to 20~TeV. The spectrum of Mkn~501 is 
evidently curved and
can be empirically fit by a power-law with an exponential cut-off at 
energy $E_0$:
$(dN_\gamma/dE) \propto E^{-\alpha}\exp(-E/E_0).$
For such a spectrum, showing a gradual steepening of the
spectral slope, one can
expect a significant spill-over of the low energy events towards the higher 
energies,
given a limited energy resolution. However, even for the rather modest energy 
resolution of
the $10\,$m Whipple telescope --- which is about $30\,\%$
\citep{mohantyetal98} --- the effect of
spill-over may not be very important, because the energy spectrum
below 10~TeV is relatively flat.
The HEGRA system of IACTs has noticeably better energy resolution (about
$10$--$20\,\%$, see \citet{hofmannetal00}). Recent studies
show that the effect of spillover for
HEGRA data is almost negligible up 17~TeV and has a limited effect above 
that energy \citep{aharonianetal99}.

In the energy range between 0.5 and 10~TeV, the Whipple Mkn~501 data
are in good  agreement with the HEGRA Mkn~501 data. Hereafter, we use in 
our analysis the
combined energy spectrum of Mkn~501 as measured by both groups, which 
extends from
260~GeV up to 22~TeV. Note that the analysis of the HEGRA Mkn~501 data 
taken during the
flaring state in 1997 has shown that the shape of the energy spectrum does 
not depend on
the flux level \citep{aharonianetal99}. This was confirmed by analysis 
of the Whipple
Mkn~501 data taken during the same observational period \citep{samuelsonetal98,
krennrichetal99}. Despite some evidence of spectral variations in Mkn~501
reported  by \citet{djannatiataietal99} and \citet{aharonianetal01a} for 
specific
observational periods, such spectral variation cannot be considered as
an established feature of the source.
Additional observations of Mkn~501 are needed in order to
prove or disprove this point. 
%{\bf 
In addition, there are no 
contemporaneous X-ray data available for those obserational periods which show  
a slight deviation of the TeV $\gamma$-ray spectra from its time-averaged spectral shape. 
Thus we do not discuss here the possibile scenario of the simulateneous X-ray and 
TeV $\gamma$-ray flare for Mkn~501, even though the main conclusions,
regarding the environmental parameters of the TeV $\gamma$-ray
emission should certainly apply to any state of Mkn~501. For the
additional fine tuning of the model parameters 
one needs more simultaneous multi-wavelengh observations of this source.
%} 

Mkn~501 is a highly variable source of TeV $\gamma$-ray emission. The
shortest variability
discovered has a doubling time of $\sim$20~mins \citep{sambrunaetal00}. 
Such fast
variability of the source is associated with sporadic changes of the 
flux level on much
longer time scales. The source can be in a flaring state ($\geq$3~Crab) 
as well as in a very
low state ($\sim$0.1 Crab) for a few months.

Occasionally Mkn~501 shows a very strong flux of X-ray emission (see
\citep{tavecchioetal01}). For example, during the 1997 flare the BeppoSAX 
satellite detected
a dramatic increase in the X-ray
flux up to 100~keV \citep{pianetal97} with a very flat spectral slope. 
Simultaneous
multi-wavelength observations of Mkn~501 revealed clear correlations between
the X-ray and the TeV
$\gamma$-ray fluxes \citep{cataneseetal97,petryetal00,sambrunaetal00}, even
though the rapid X-ray flares from Mkn~501 are not always accompanied by 
a high emission
state in TeV $\gamma$-rays \citep{catanesesambruna00}.

%{\bf 
We used here the data taken with {\it BeppoSAX} in observation period of April 1997 
as discussed in detail by \citep{pianetal97}. These data were combined 
with the time-averaged TeV $\gamma$-ray spectrum as measured by HEGRA in 1997 observational 
campaign \citep{aharonianetal99}.
%} 

\subsection{Mkn~421}

The BL~Lac object Mkn~421 is one of the AGNs detected by the EGRET 
instrument on board
Compton GRO in the $30\,$MeV--$30\,$GeV energy range \citep{thompsonetal95}. 
It was the first
extra-galactic TeV $\gamma$-ray source, discovered by the Whipple group 
\citep{punchetal92}
and was confirmed by the HEGRA group \citep{petryetal96}. Since its 
discovery, Mkn~421 has
shown a very low baseline TeV $\gamma$-ray emission with a few extremely 
rapid flares
on timescales from one day to 30 minutes \citep{gaidosetal96}. The limited 
statistics
of the $\gamma$-ray events recorded during the short duration flares, 
as well as the long
exposure observations during the very low emission state, did not permit a
measurement of the energy
spectrum above $7\,$TeV \citep{krennrichetal99,aharonianetal99a}. 
Only recently, in
2000 and 2001, has Mkn~421 demonstrated a flare with an average flux of 4~Crab

\citep{krennrichetal01,aharonianetal02,krennrichetal02}. Data taken during 
this flare have been used
to extract the energy spectrum of Mkn~421 at higher energies, up to 20~TeV. 
As stated
by both groups, the energy spectrum of Mkn~421 is evidently curved. Analysis of
the HEGRA Mkn~421 data revealed significant variations of the spectral 
slope at energies below 3~TeV
\citep{aharonianetal02,krennrichetal02}, with the high flux state 
showing a substantially harder spectrum. However the
best empirical fit to the data taken in both the high and the low states 
gives the same cut-off energy of
3.6~TeV. This is consistent with the assumption that all spectra 
are affected by
intergalactic absorption.
The cut-off energy derived from the Whipple data on Mkn~421 
\citep{krennrichetal01,krennrichetal02}
is $4\,$TeV. Given the statistical and systematic errors of such a measurement
this value agrees
with that measured by HEGRA group \citep{aharonianetal02}.

Mkn~421 is a bright, variable X-ray source. Its X-ray energy spectrum 
was measured in different
states of emission by the BeppoSAX instrument \citep{fossatietal00} 
within the range from 0.1~keV
up to 10~keV and the RXTE instrument up to 100~keV. During the 
flares, the peak of
the X-ray emission
moves towards high energies \citep{fossatietal00}. Note that the peak 
position in the X-ray
spectrum of Mkn~421 is substantially lower ($\sim$1 keV) than the peak 
in the X-ray
spectrum of Mkn~501 ($\sim$100 keV) in its flaring state. This implies
intrinsic physical differences between the two sources.

%{\bf 
We used here the fits of the X-ray spectra of Mkn~421 measured with the 
{\it BeppoSAX} in 2000 for low and high states of the source 
\citep{fossatietal02} (see Figure~5). These data were combined with the two time 
averaged HEGRA spectra \citep{aharonianetal02} of Mkn~421 derived from the data taken 
mostly in 2000 (see Figure~2 in \citep{aharonianetal02}) for two bservational periods 
when Mkn~421 was in high and low states.
%}   

\subsection{1ES~1426+428}

1ES~1426+428 is a BL~Lacertae object (see \citep{remillardetal89}) 
at a redshift of $z=0.129$.
Recent BeppoSAX observations \citep{costamanteetal01} revealed for 
this source a rather flat ($\alpha = 0.92$)
power-law X-ray spectrum extending up to 100~keV, without evidence 
for a spectral break.
Such extreme behavior of the source in X-rays is reminiscent of the BL~Lacertae
object Mkn~501 previously detected at TeV energies. Estimates of 
the TeV $\gamma$-ray
flux from 1ES~1426+428 obtained in a homogeneous synchrotron 
self-Compton (SSC) model
\citep{costamanteetal01} suggested that this source should be 
detectable at TeV energies.

This was confirmed by the detection of 1ES~1426+428 recently reported by
the Whipple group \citep{horanetal02}. The source was also detected by 
the HEGRA group \citep{aharonianetal02b} during
the same observational period in 2000 and 2001. HEGRA also measured 
the energy spectrum of
1ES~1426+428 \citep{aharonianetal02b}. Recent spectral measurements made by 
Whipple \citep{petryetal02} and CAT \citep{djannatiataietal02} 
generally confirmed HEGRA spectrum of 1ES~1426+428.
The limited statistics of the detected $\gamma$-rays
does not allow a study of flux and spectral variability. A preliminary estimate
of the spectral shape is consistent with a rather steep power-law 
spectrum and a significant
$\gamma$-ray rate above 3~TeV \citep{aharonianetal02b}. Note that 1ES~1426+428
has a redshift substantially
larger than the two well-established BL~Lac objects, Mkn~421 and Mkn~501.
Therefore, IR absorption should significantly modulate the intrinsic TeV
$\gamma$-ray spectrum of this object. However, current 1ES~1426+428 
data are rather limited
in statistics and cannot be used to place stringent constraints 
on the distribution of EBL.

\subsection{Other TeV BL Lacs}

For the three remaining AGN detected at TeV energies, namely 
1ES~2344+514 \citep{cataneseetal98}, 
1ES~1959+650 \citep{nishiyamaetal00}, and PKS~2155-304 \citep{chadwicketal99},
there are still no data available on the spectral shape. 1ES~1959+650
was detected recently by Whipple \citep{weekesetal02} and HEGRA 
\citep{hornskonopelko02} in a
very high state ($\geq3$~Crab), but the spectral measurements have not yet
been published.
The TeV $\gamma$-ray fluxes reported for 1ES~2344+514, 1ES~1959+650, 
and PKS~2155-304,
are consistent with the IR absorption model used here. These data do not
constrain the model severely because of the rather large uncertainties 
of the measured TeV $\gamma$-ray spectra.

\clearpage
\begin{figure}
\plotone{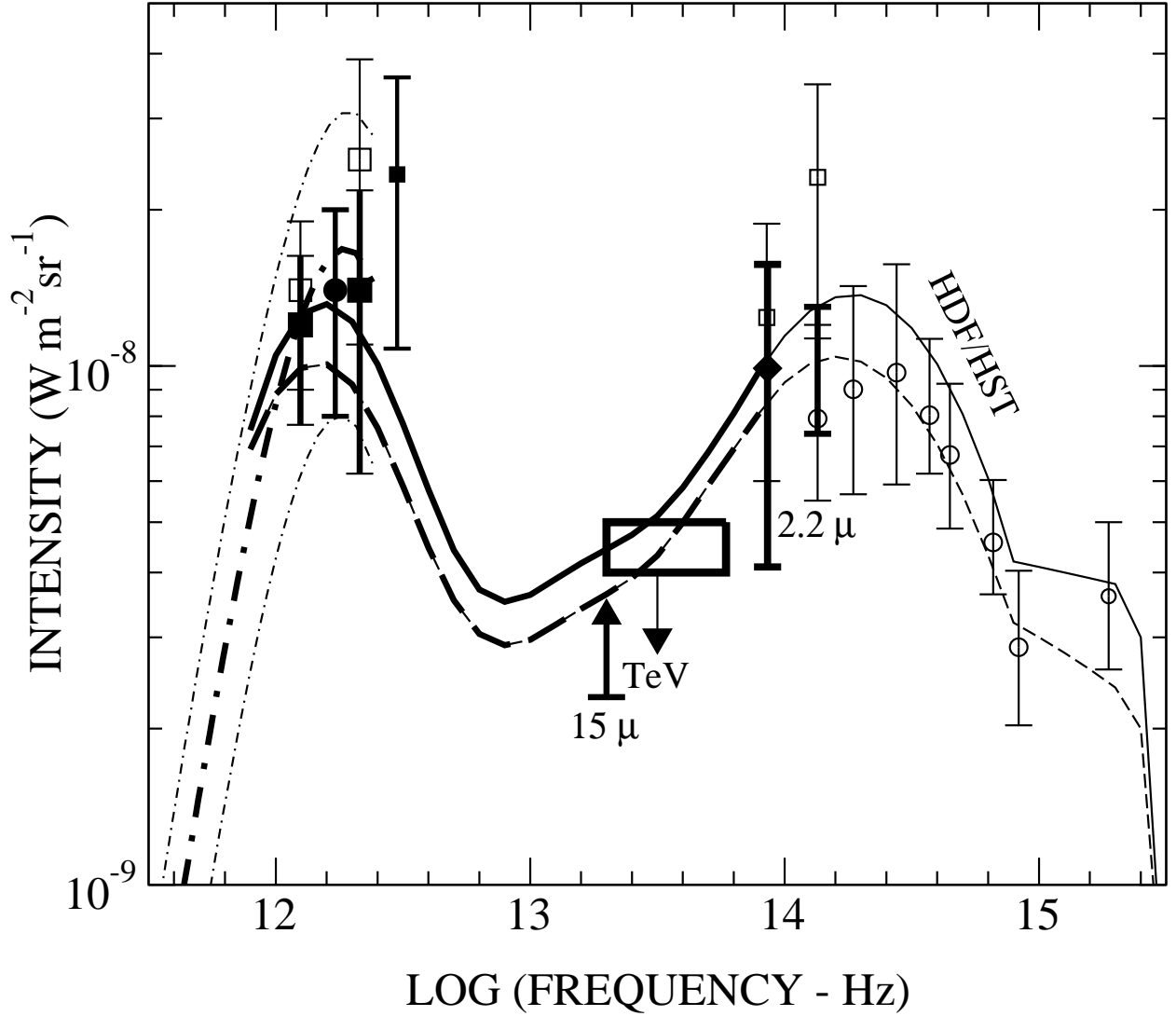}
\caption{The SED of the EBL. Solid and dashed curved correspond to the
  \citet{malkanstecker01} data obtained assuming the {\it rapid} as well as
  {\it baseline} luminosity evolutionary model, respectively. The
  infrared luminosities of galaxies is assumed to increase as
  $(1+z)^3$ from $z_1=0$
  back to $z_2=2$ for the lower curve, and as $(1+z)^4$ from $z=0$ back to
  $z_2=1.4$ for the upper curve. Both curves have been derived assuming a
  flat star formation rate at $z>z_2$. \label{absorption}}
\end{figure}

\clearpage

\section{Opacity of TeV $\gamma$-rays}
TeV $\gamma$-rays can be absorbed and produce
electron-positron pairs on interaction with photons of the
intergalactic background light (see \citet{steckerdejagersalamon92}).
The corresponding opacity can be calculated once an assumption is made about
the spectral energy
distribution (SED) of the extra-galactic background light (EBL). $\gamma$-rays
of energy from 300~GeV to 20~TeV interact most effectively with photons of
wavelength from $1$--$50\microm$, respectively. There have been a number of
attempts in past to measure the SED of EBL in this range
as well as to model it (for a review see \citet{hauserdwek01}).
Recently \citet{malkanstecker01}
revised their empirically based model \citep{malkanstecker98} for the EBL
starting from the near-IR and extending to the far-IR range
(see Fig.~\ref{absorption}). This model is consistent with
all currently measured lower and upper limits in the mid-IR range 
as well as with the
COBE-FIRAS fluxes given at $140\microm$ and $240\microm$. In addition, 
\citet{dejagerstecker02} extended the EBL derived in 
\citep{malkanstecker01} to ultraviolet (UV) wavelengths
using the galaxy counts in the Hubble Deep Field data taken with 
Hubble Space Telescope
(see \citet{madaupozzetti00}). The so-called {\it hybrid} model developed by
\citet{dejagerstecker02}
gives a smooth distribution of the EBL over the $1$--$300\microm$
wavelength range. It is consistent with existing measurements
and upper limits in the optical--UV range and agrees rather well with similar
theoretical models (e.g. \citet{tansilkballand99,rowanrobinson01,xu00})
in the mid-IR range. Note that all of those models predict a 
flat distribution of EBL
in the mid-IR range at a flux level of about 
$3$--$4\,{\rm nW\,m^{-2}sr^{-1}}$. However,
direct measurements of the EBL in the mid-IR range are hampered by unavoidable
foreground emission (for a detailed discussion see \citet{hauserdwek01}).

An alternative approach to finding the SED of EBL in the relevant energy range
is provided by semi-analytical modelling of the evolution of galaxy formation
(see \citet{primack02}). However, the results of this approach 
are sensitive, for
example, to the IMF assumed in
the calculations. In fact different IMFs result in rather different
SEDs (see \citet{primack02} and references therein).
Specific sets of model parameters succeed rather
well in
reproducing the phenomenological hybrid model developed by 
\citet{malkanstecker01},
and such models could in principle be used in conjunction with an SSC model of
the intrinsic emission of an AGN to predict the TeV flux.
However, in this
paper it is not our goal to test both the SSC and the absorption models
simultaneously. Instead, we wish to test only the SSC model
and include the uncertainties in the level of absorption into
an additional error to be added to that already present in the TeV
observation.
To this end, we consider the two extreme cases of
high and low flux SEDs as given by \citet{malkanstecker01} as shown in
Fig.~\ref{absorption}.
One can see that in the mid-IR region the
upper limits derived from the TeV $\gamma$-ray data  
(see e.g. \citet{renaultetal01}) and
the low limits established from the data taken by ISOCAM instrument (see 
\citet{franceschinietal01}) are very close to each other. 
They stringently limit the allowed region for the actual
SED of EBL in this range.
For the extreme cases considered by \citet{malkanstecker01}, 
\citet{dejagerstecker02} have calculated the
optical depth, $\tau(E_\gamma,z)$, of the $\gamma$-rays of energy $E>50$ GeV
for
redshifts up to $z=0.3$.
Using the parametric expression given in Eq.~(1,2) in \citet{dejagerstecker02},
we interpret the resulting upper and lower values of the \lq\lq
de-absorbed\rq\rq\ intrinsic flux
\begin{equation}
(dN_\gamma/dE)_{\rm intrinsic}=(dN_\gamma/dE)_{\rm measured} 
\cdot \exp\left[\tau (E,z)\right].
\end{equation}
as
$3\sigma$ deviations from the true value, and add this error to that
attached to the original TeV point.

The resulting intrinsic spectra for both the flaring
and quiescent states of Mkn~501
and for Mkn~421 shown in Figure \ref{mkn501fit},\ref{mkn421fit}.
All of these spectra show a prominent peak in the
$E^2 F(E)$ distribution The position of the peak is at about 
8~TeV and 2~TeV for Mkn~501 and Mkn~421,
respectively.
These objects have significantly different X-ray spectra, which are shown in
Figs.~\ref{mkn501xrays} and \ref{mkn421xrays}, respectively.

\section{Modelling}

To model the multi-wavelength spectra of the BL~Lac objects Mkn~421 and
Mkn~501 in a
homogeneous SSC scenario we use an approach described by
\citet{mastichiadiskirk95,mastichiadiskirk97}. This method involves
prescribing an injection function for relativistic electrons and solving the
two
time-dependent kinetic equations for the electron and photon distributions
source.
All relevant physical processes are taken into account in the code,
i.e., synchrotron radiation, inverse Compton scattering 
(both in the Thomson and
Klein-Nishina regimes), photon-photon pair production, and 
synchrotron self-absorption.
Both synchrotron and inverse Compton scattering emissivities have been
improved
from the original version \citep{mastichiadiskirk95} and now incorporate
the full
emissivity rather than a delta-function approximation.

Seven model parameters are required to specify a source in a stationary
state. These are:
\begin{enumerate}
\item
the Doppler factor $\doppler=1/[\Gamma(1-
\beta\cos\theta)]$, where $\Gamma$ and $c\beta$ are the Lorentz factor
and speed of the source, and $\theta$ is the angle between its direction
of motion and the line of sight to the observer,
\item
the radius $R$ of
the source (in its rest frame, in which it is assumed spherical)
or, equivalently, the crossing
time $\tcross=R/c$. This is related to the observed minimum
variation timescale
in the galaxy frame by $\tvar=R/(\doppler c)$,
\item
the magnetic
field strength $B$,
\item
the index $s$ of the electron injection spectrum, for which we take
$\elinject=\lowq \gamma^{-s}e^{-\gamma/\gammamax}$ where $\gamma$ is the
electron Lorentz factor,
\item
$\gammamax$, the Lorentz factor at the cut-off of the injection spectrum
\item
$\lowq$, the amplitude of the injection spectrum, 
which is expressed in terms of the electron
injection compactness $\elcomp={{1\over 3}\melec c\sigmaT R^2}
\int_1^\infty {\rm d}\gamma (\gamma-1) Q_{\rm e}$
\citep{mastichiadiskirk95},
\item
$\tesc$, the effective escape time
of relativistic electrons, which can be identified as the timescale over
which adiabatic expansion losses limit the accumulation of relativistic
electrons within the source.
\end{enumerate}

It is possible to introduce a lower limit to the
Lorentz factor at which electrons are injected \citep{krawcoppiaharonian02}. 
This permits more flexibility in fitting the soft X-ray
spectrum and leads to different values of the parameter $s$, which is
then no
longer related to the low frequency spectral index. However, at TeV
$\gamma$-ray energies, the models are essentially identical.

In attempting to optimize a fit to a
particular data set, it is essential to use a
physically motivated strategy to arrive at reasonable starting values for
these seven parameters. This is done by identifying six scalars which
characterise the typical blazar spectrum:
\begin{enumerate}
\item
the peak frequency $\nusync$
of the synchrotron emission expressed in units of $10^{18}$Hz
\item
\label{icpeak}
the peak frequency $\nucompt$ of the inverse Compton emission, expressed
in units of $10^{27}$Hz
\item
the total nonthermal luminosity $L$
\item
\label{ratio}
The approximate ratio $\eta$
of the total flux in the inverse
Compton part to that in the synchrotron part of the spectrum
\item
the break frequency $\nubreak$ in the synchrotron part
(typically between millimeter and optical wavelengths)
of the spectrum, where
cooling and electron escape or expansion losses are comparable
\item
the low-frequency spectral index
\end{enumerate}

Together with an
estimate of the fastest variability timescale $\tvar$,
these roughly estimated quantities or {\em observables}
enable one to find reasonable starting values of the
seven parameters of the SSC model.
However, this method works
only if the intrinsic spectrum is available. In particular, it is not possible
to estimate quantities $\nucompt$ (\ref{icpeak}) or $\eta$ (\ref{ratio})
unless the observed spectrum has been
unfolded using an absorption model.

Following \citep{mastichiadiskirk97} we can find an estimate
of the Doppler factor in terms of the observables:
\eqb
\doppler&=&55\, \nucompt^{1/2}\,\tvarthree^{-1/4}\,\nusync^{-1/4}\,
L_{46}^{1/8}\,\eta^{-1/8}
\label{dopplerf}
\eqe
where $L_{46}$ is the nonthermal luminosity in units of $10^{46}$erg/s and
$\tvarthree$ is the variation timescale in units of $10^3$s. From this follow
the radius, magnetic field and Lorentz factor at the cut-off:
\eqb
R&=&c\tvar\doppler;\quad  B\,=\,5\times10^{-3}\doppler\nusync\nucompt^{-2}{\rm
  G};
\quad \gammamax\,=\,3\times10^6\nucompt\doppler^{-1}
\label{therest}
\eqe
The power-law index $s$ is limited by radio data of Mkn~421 and Mkn~501 to be
close to $1.5$--$1.7$
provided, as we assume here, that the radiowaves
are created in the same region as the high energy photons.
In this case, the parameter $\gammamax$ is important,
since most of the power is injected at the highest permitted values of the
Lorentz factor.
The remaining parameters are the electron compactness $\elcomp$,
and the escape time $\tesc$ which do not have a strong influence on the
shape of the X-ray and TeV spectra, but only on the flux levels.

%{\bf
Using these starting values, we find the best fit models for
the high and low states of Mkn~421 and Mkn~501 as
given in Table~\ref{ssctable}. In each case, the differences between the
optimal parameter values and the starting values given by Eq.~(\ref{dopplerf})
and (\ref{therest}) are small. The $\chi^2$ values associated with these fits,
taking into account the error introduced by the uncertainty in intergalactic
absorption
are unusually small. This indicates two things: 1) the SSC model provides a
good fit and 2) the errors inferred from uncertainty in SED of IR   
are probably too generous. 
We may conclude that the actual value of the intergalactic absorption lies 
closer
to the mean of the two extreme models given by \citet{malkanstecker01}
than
suggested by our assumption that these represent  $1\sigma$ deviations.
%}

\begin{figure}
\plotone{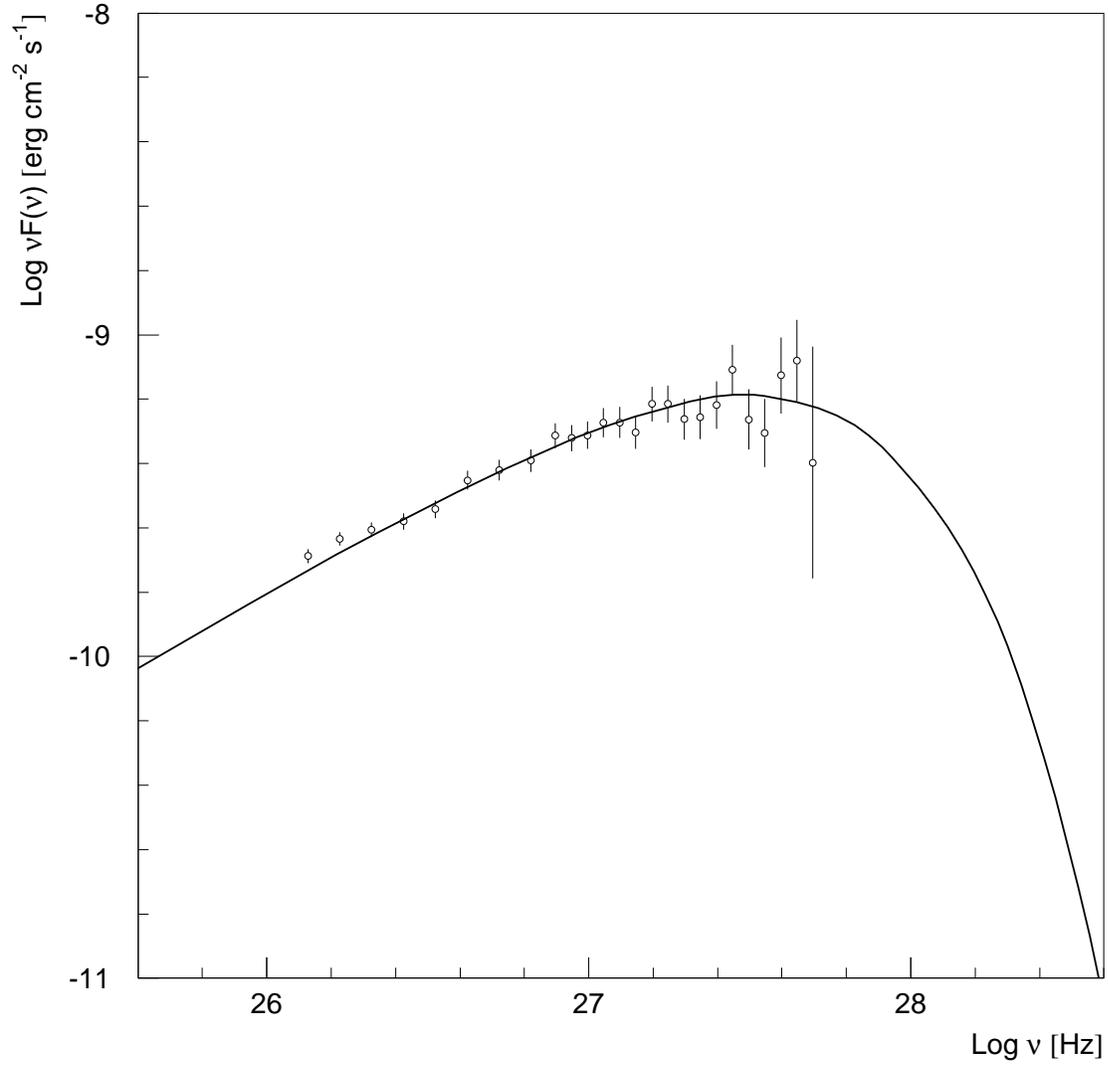}
\caption{\label{mkn501fit}
The de-absorbed spectrum of Mkn~501, with error bars indicating the
uncertainty in intergalactic absorption, together with the best fit SSC model
}
\end{figure}

\begin{figure}
\plotone{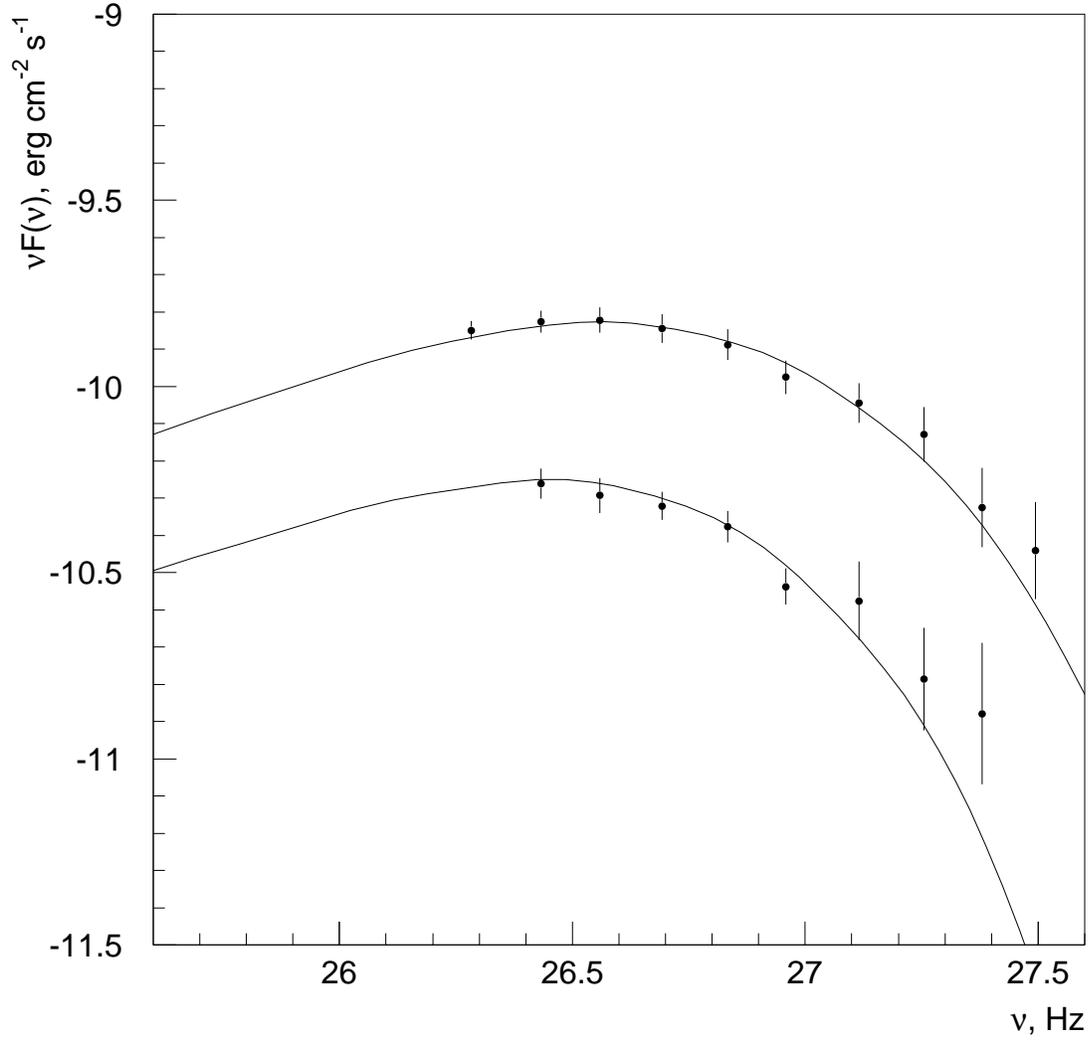}
\caption{\label{mkn421fit}
The de-absorbed spectrum of Mkn~421, with error bars indicating the
uncertainty in intergalactic absorption, together with the best fit SSC models
for both the high and low states
}
\end{figure}

\begin{figure}
\plotone{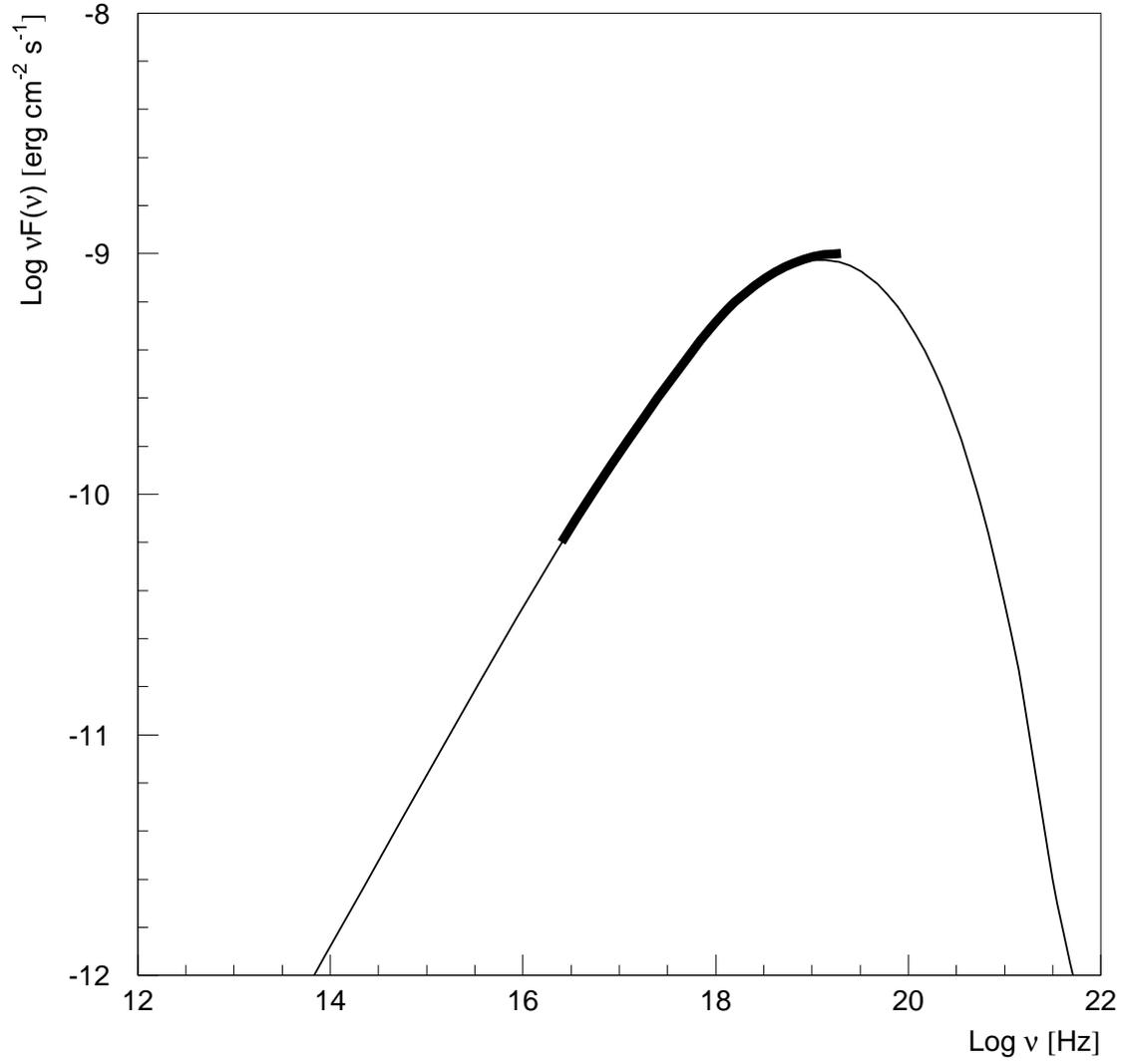}
\caption{\label{mkn501xrays}
The X-ray spectrum of Mkn~501, 
together with the best fit SSC model
}
\end{figure}

\begin{figure}
\plotone{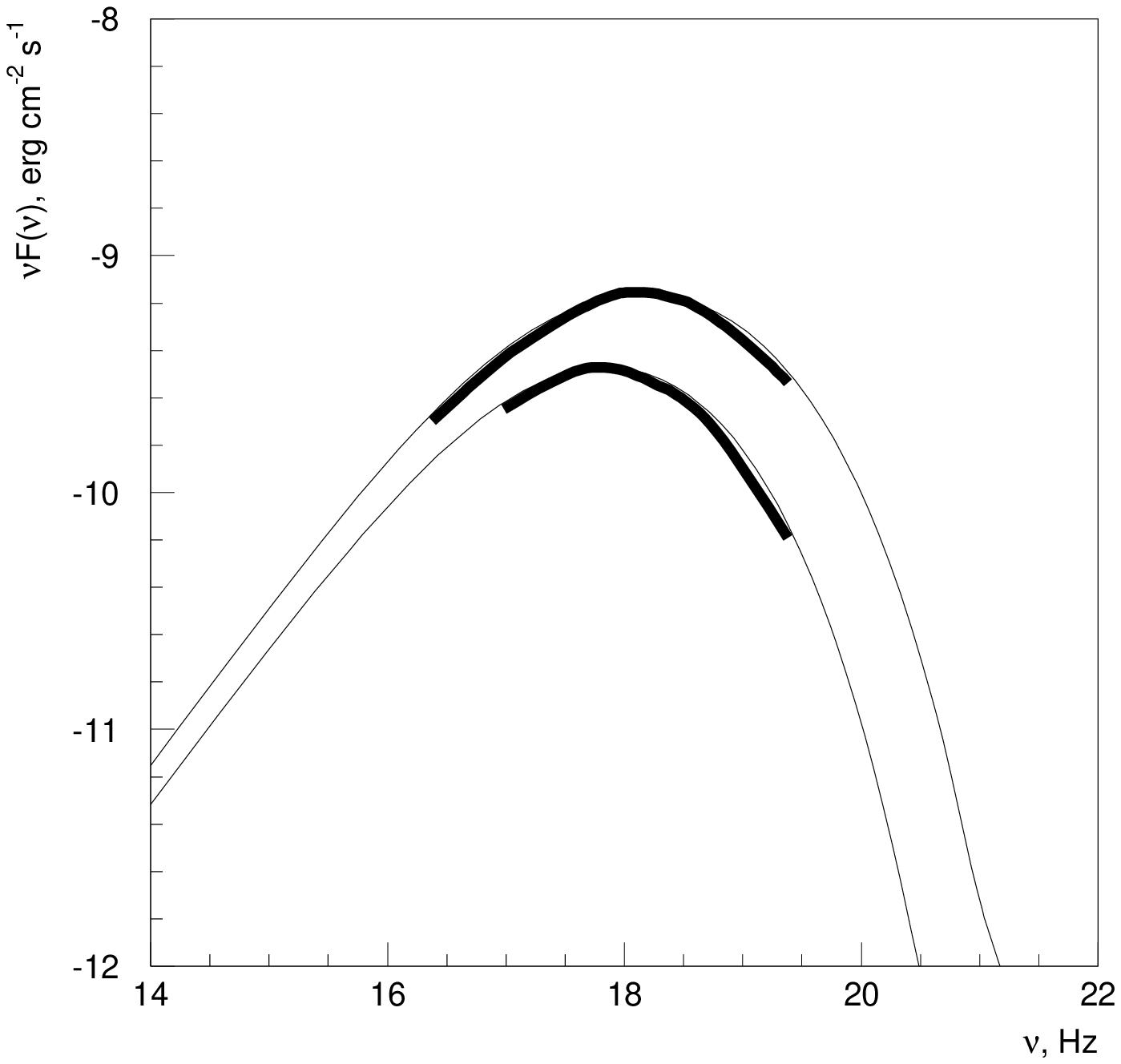}
\caption{\label{mkn421xrays}
The X-ray spectrum of Mkn~421, 
together with the best fit SSC models
for both the high and low states
}
\end{figure}

In Figure \ref{mkn501fit} and \ref{mkn421fit}
we show the fits for both BL~Lac objects, Mkn~501 and
Mkn~421 in the TeV range, in the case of Mkn~421 in both high and low states.
In order to derive the time averaged spectrum
of Mkn~501 (see also Section~2),
which extends up to 17~TeV, we
assume that despite
the different levels of emission,
the TeV $\gamma$-ray {\em spectrum} of Mkn~501 remains the same as that
derived from the HEGRA data \citep{aharonianetal99}.\footnote{CAT group claimed
an evidence for spectral variability in Mkn~501 $\gamma$-ray spectrum during
the same observational period as HEGRA \citep{djannatiataietal99}. 
However CAT instrument has substantially
lower energy threshold, $E_{th}\simeq 250$~GeV,
whereas HEGRA has energy threshold
of about 500~GeV. Here we are interested in data around and above 1~TeV, where
spectra taken at different fluxes have similar spectral shape.}
The same models are shown together with the X-ray data in
Figs.~\ref{mkn501xrays} and \ref{mkn421xrays}. 
%{\bf 
The complete SSC fits for Mkn~501 and Mkn~421 are shown in Figure~6 and 7.
%}

Interestingly, the $\gammamax$ values (see Table~\ref{ssctable}) 
providing the best fit to
the data are very different for Mkn~421 and Mkn~501. The position of a peak in
the de-absorbed $\gamma$-ray spectrum of Mkn~501 is at about 8~TeV, whereas for
Mkn~421 it is at noticeably lower energy of about 2~TeV. The value of 
$\gamma_{max}$
strongly correlates with the maximum energy of the observed IC photons 
as given by
e.g. \citet{kinotakaharakusunose02}.
We also find that the main difference between the high and low
states of Mkn~421 can be accounted for by a change in $\gamma_{max}$ 
%{\bf 
(as discussed 
earlier by \citep{fossatietal00})
%}
and in luminosity 
which, however, is not large. 

Recent observations of Mkn~421 and
Mkn~501, have revealed variability on the very short time scale of a few times
$\sim10^3$s (see Section~2). Combined with the fact that the de-absorbed
spectrum of Mkn~501 is rather flat at TeV energies \citep{konopelkoetal99}
 (so that $\nucompt>1$), it is
evident from Eq.~(\ref{dopplerf}) that the high Lorentz factors $\sim50$ are
required to fit the observations, since this quantity is very insensitive to
the other properties of the spectrum.
This is in agreement with earlier estimates \citep{mastichiadiskirk97,
konopelkoetal99}.

In order to check that the
choice of the parameters does indeed lead to variations on the timescale
$\tvar$
in the TeV regime, we performed time-dependent
calculations.
Starting from a stationary state,
we impulsively
changed the electron injection compactness by a factor of several
and checked the behaviour of the flux at the TeV regime as it moved
towards the new stationary state. We found that for all cases
prescribed in Table~\ref{ssctable},
the e-folding time of the flux was always between
$\tvar$ and $2\tvar$.

\clearpage
\begin{deluxetable}{llllllllll}
\tabletypesize{\scriptsize}
\tablecaption{Summary of physical parameters used for the SSC models. \label{ssctable}}
\tablewidth{0pt}
\tablehead{
\colhead{Source} & \colhead{State}   &
\colhead{$\delta$} & \colhead{$R$ (cm)}   &
\colhead{$s$} & \colhead{$\gammamax$} & \colhead{$l_e$} &
\colhead{$B$ (G)}     & \colhead{$t_{cros}/t_{esc}$} & 
\colhead{$\chi^2/dof$}
}
\startdata
Mkn~501 & -    & 50 & $1.5 \times 10^{15}$ & 1.55 & $1.0 \times 10^6$ & $6.20 \times 10^{-5}$ & 0.10 & 1.0 & 0.86 \\ \hline
Mkn~421 & high & 55 & $1.5 \times 10^{15}$ & 1.65 & $3.0 \times 10^5$ & $1.74 \times 10^{-5}$ & 0.45 & 1.0 & 0.33 \\
Mkn~421 & low  & 55 & $1.5 \times 10^{15}$ & 1.65 & $1.7 \times 10^5$ & $1.15 \times 10^{-5}$ & 0.40 & 1.0 & 1.32 \\
\enddata
\end{deluxetable}

\clearpage

It is generally accepted that the X-ray fluxes from BL~Lac
objects like Mkn~421 and Mkn~501, are well correlated with the TeV
$\gamma$-ray fluxes --- the high state emission in X-rays
appears at the same time in the TeV $\gamma$-rays. BL~Lac objects
can stay in a high emission state for more than 6 months (e.g., Mkn~501)
and can at the same time be highly variable sources of both X-rays
and $\gamma$-rays on a time scale of less than one hour. The spectral
variations in TeV $\gamma$-rays are not well studied so far at such short
time scales, even though there are clear indications for such variations
(at least in case of Mkn~421, see \citet{aharonianetal02}).

The behavior of the X-ray emission of the BL Lac objects is sporadic, and
also demonstrates very large flux variations on extremely short time scales of
less than one hour \citep{sambrunaetal00,catanesesambruna00}.
TeV $\gamma$-ray flares on similar time scales usually do not permit the
accumulation of a sufficient number of $\gamma$-ray events with currently
operating ground based Cherenkov detectors. Poor event statistics in such
cases prevent the measurement of the energy spectrum well above 3~TeV, and
make it difficult to detect the effects of IR absorption in such spectra.
Thus, we feel it is premature to study in detail the correlations
between the X-ray and multi-TeV $\gamma$-ray emission spectra on 
such short time
scales, and have restricted our modelling to time-independent states.

\clearpage

\clearpage

\clearpage

For Mkn~421 a variation of
the spectral slope with flux level was recently discovered
by HEGRA \citep{aharonianetal02}. However here we are not aiming to fit
the X-ray and TeV $\gamma$-ray spectra for each particular flare,
because the observational data in X-rays and TeV $\gamma$-rays
are insufficient. Thus we consider here
only two different spectra of X-ray and TeV $\gamma$-ray emission 
from Mkn~421 in lower
and high state. It is important to mention that the position of the cut-off
energy is the same in each case ($E_0 =3.6$~TeV).

\clearpage

\clearpage

\clearpage

\begin{figure}
\plotone{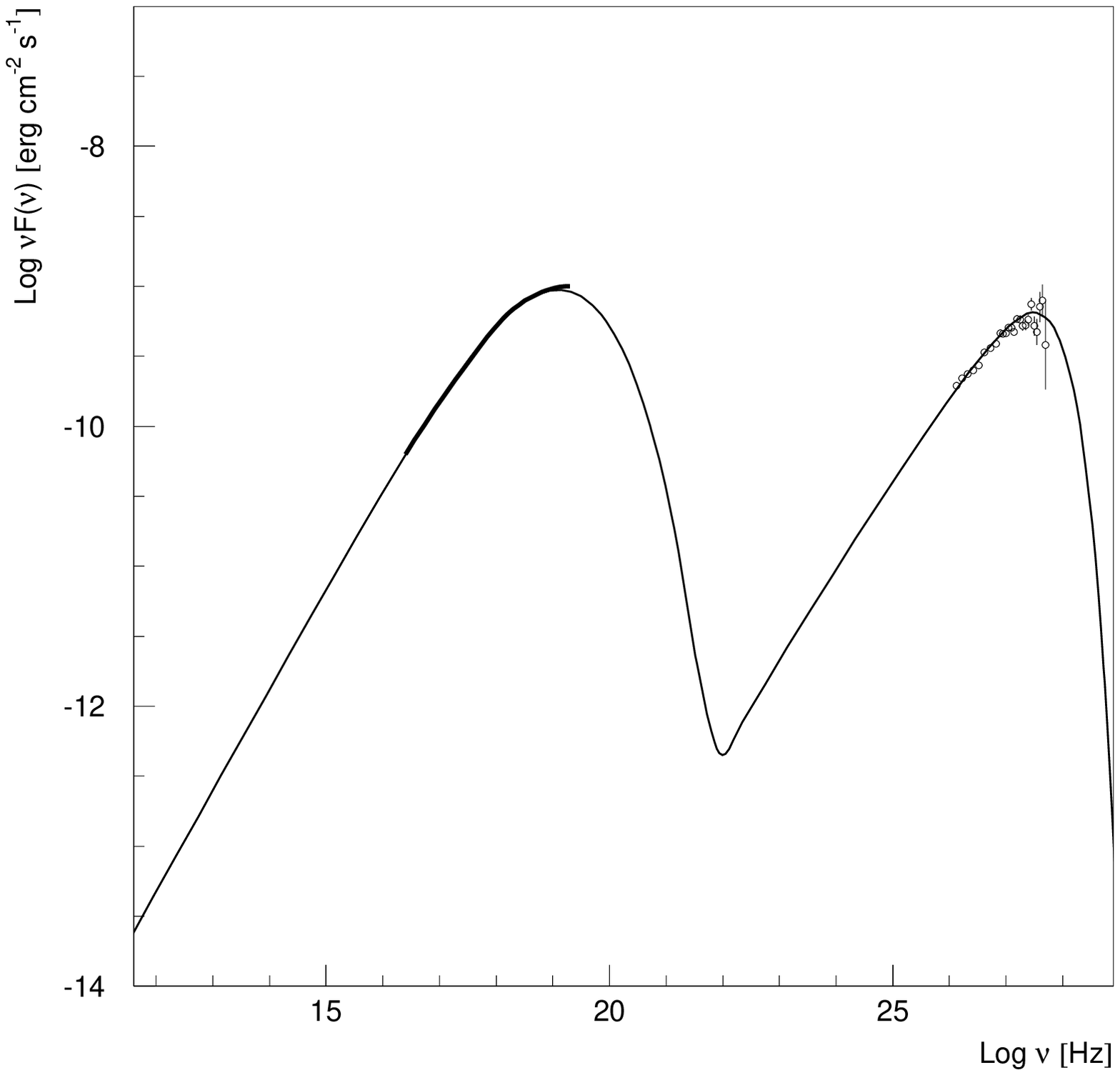}
\caption{\label{mkn501all}
The combined X-ray/TeV $\gamma$-ray spectrum of Mkn~501 together with 
the best fit SSC model.}
\end{figure}

\clearpage

\begin{figure}
\plotone{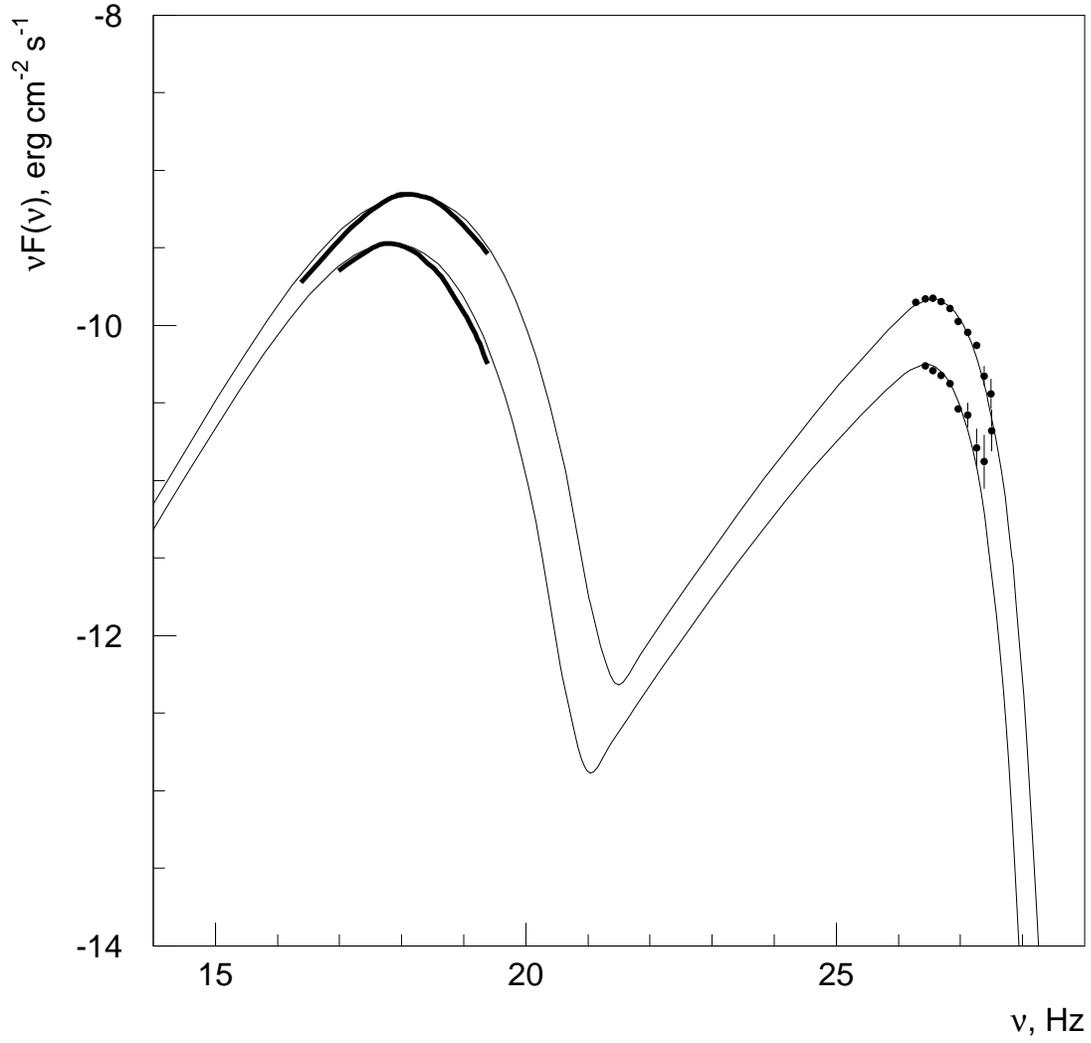}
\caption{\label{mkn421all}
The combined X-ray/TeV $\gamma$-ray spectrum of Mkn~421,
together with the best fit SSC models
for both the high and low states.
}

\end{figure}

Figure \ref{mkn501xrays} shows
BeppoSAX measurements
taken exactly 
during the 1997 outburst of Mkn~501. The X-ray and TeV
$\gamma$-ray spectrum were averaged over different observing times but
their fluxes at long time scales (month) are strongly correlated and finally
X-ray to TeV $\gamma$-ray flux ratio remains almost unchanged.

\section{Conclusion}

We use here the data on the energy spectra of Mkn~501 and Mkn~421 over the 
energy range from 500~GeV up to $\simeq$20~TeV, which 
became available recently, to reconstruct the intrinsic, 
IR de-absorbed source spectra for
both of these AGNs. Present uncertainties in EBL are treated here as the 
systematic errors of the de-absorbed spectra. We fit those intrinsic spectra 
of Mkn~501 and Mkn~421 (in low and high states) along with the X-ray data 
using the homogeneous SSC model.
 
Both the X-ray data and de-absorbed TeV $\gamma$-ray data can be fitted
reasonably well by a homogeneous SSC model, even though the X-ray
and TeV $\gamma$-ray features of both Mkn's are very different.

The two AGNs modelled
   have definitely different intrinsic spectra as well as slightly
   different variability scales. This leads to
    a slightly higher value of Doppler factor for Mkn~421
as well as a higher magnetic field and a lower value of
$\gammamax$.

The energy spectra of Mkn~421 in high and low states can be fitted by
   changing the $\gammamax$ and luminosity parameters,
   whereas the Doppler factor and
   the magnetic field remain unchanged.

%{\bf 
In present analysis we used the X-ray and TeV $\gamma$-ray data, 
which are not simulateneous. They were averaged over not exactly the 
same but overlapping observing periods. This is because, first of all,  
one needs a substantially longer period to observe in TeV $\gamma$-rays 
and measure the spectrum over a broad multi-TeV region. Secondly, the 
shape of TeV $\gamma$-ray spectrum is more constraining than the absolute 
flux level. However, further dedicated multiwave
campaigns for Mkn~421 and Mkn~501 may place a tighter limit on the choice 
of the model parameters.
%}   

We argue that a logically consistent SSC model of X-ray and TeV $\gamma$-ray 
emission can be constructed for both AGNs, Mkn~501 and Mkn~421, taking into 
account the IR absorption. Accurate spectral data are needed for 
other AGNs, in particular for the newly discovered 
1ES~1426+428 and 1ES~1959+650, in order to test the consistency of this 
model for AGNs at different redshits.

\end{document}